# Data centre optical wireless downlink with WDM and multi-access point support

Osama Zwaid Alsulami[1], Sarah O. M. Saeed[1], Sanaa Hamid Mohamed[1],
T. E. H. El-Gorashi[1], Mohammed T. Alresheedi[2] and Jaafar M. H. Elmirghani[1]
[1]School of Electronic and Electrical Engineering, University of Leeds, LS2 9JT, United Kingdom
[2]Department of Electrical Engineering, King Saud University, Riyadh, Kingdom of Saudi Arabia
ml15ozma@leeds.ac.uk, elsoms@leeds.ac.uk, elshm@leeds.ac.uk,
t.e.h.elgorashi@leeds.ac.uk, malresheedi@ksu.edu.sa, j.m.h.elmirghani@leeds.ac.uk

**ABSTRACT**
The ability to provide very high data rates is a significant benefit of optical wireless communication (OWC) systems. In this paper, an optical wireless downlink in a data centre that uses wavelength division multiple access (WDMA) is designed. Red, yellow, green and blue (RYGB) laser diodes (LDs) are used as transmitters to provide a high modulation bandwidth. A WDMA scheme based on RYGB LDs is used to provide communication for multiple racks at the same time from the same light unit. Two types of optical receivers are examined in this study; an angle diversity receiver (ADR) with three branches and a 10 pixel imaging receiver (ImR). The proposed data centre achieves high data rates with a higher signal-to-interference-plus-noise ratio (SINR) for each rack while using simple on-off-keying (OOK) modulation.

## 1. INTRODUCTION

Optical wireless communication (OWC) systems can be attractive in data centres due to their energy efficiency, scalability and flexibility. Thus, hundreds of metres of cables or optical fibre are replaced by using OWC systems. A visible light communication (VLC) system is a type of OWC system that can be used for downlink communication in the data centre. VLC is a promising technology that can provide high data rates and has been a matter of interest for researchers interested in data transfer [1]–[8]. The radio frequency (RF) spectrum is becoming scarce as the demand for higher data rates is growing. Offering very high data rates that exceed 10 Gbps, sometimes reaching the Tbps region, by using radio systems is highly challenging. Cisco states that, Internet traffic is expected to grow 27 times between 2016 and 2021 [9]. It has been shown by many studies that VLC systems can transmit video, data and voice contents at high data rates, up to 20 Gbps in indoor environments [7], [8], [18], [19], [10]–[17], [20]. In addition, VLC can offer a license-free bandwidth that has high security and is low-cost compared to RF systems [12], [21]–[28]. However, some limitations of the VLC system have been demonstrated, such as the absence of line-of-sight (LOS) components in some links which significantly reduce the system's performance. Also, inter-symbol interference (ISI), caused by the multipath propagation, can reduce the system's performance. In addition, interference from different users can also have an impact on the performance of the system. OWC multiple access techniques can be used to reduced interference from other users by employing a variety of orthogonal coding methods, such as a wavelength division multiple access (WDMA). Recently, WDMA has received attention, especially in the context of supporting multiple users [4], [29]–[33].

This paper proposes a data centre based on a downlink VLC system. A WDMA scheme based on red, yellow, green, and blue (RYGB) Laser Diodes (LDs) is used in this work in conjunction with two types of receivers; an angle diversity receiver (ADR) and an imaging receiver (ImR). The effects of multipath propagation are considered in this work. White colour can be provided by using RYGB LDs for indoor illumination, as reported in [19]. In addition, RYGB LDs can be used as transmitters to provide a high modulation bandwidth. The rest of the paper is organised as follows: The data centre configuration is described in Section 2 and the optical receiver design is discussed in Section 3. Section 4 introduces the optical transmitter design and Section 5 shows the simulation results, and the conclusions are provided in Section 6.

## 2. DATA CENTRE CONFIGURATION

The dimensions of a pod in the data centre are assumed to be (length × width × height) 8 m × 8 m × 3 m in this study, similar to [8]. Three rows of racks are assumed in the data centre and each row consists of 10 racks [34]–[37]. Each rack has its own top of rack (ToR) switch which is placed at the top of the rack and uses the red colour, as shown in Figure 1b. The ToR switch is used as a communication coordinator between servers inside the rack and the rest of the data centre. The rack dimensions (length × width × height) are shown in Figure 1b. In the proposed data centre, more than one meter has been set between rows of racks, and between each row of racks and walls for ventilation. In the simulation, a ray-tracing algorithm was utilised for modelling the ceiling, walls and floor reflections inside the data centre [38]. Each surface inside the data centre was divided into small equal areas ($dA$) with a reflection coefficient ($\rho$). Reflections up to second-order were considered in the simulation due to the fact that reflections higher than the second-order have no effect on the received signal [38]. It has been shown that plaster walls which are considered in this work reflect signals in a Lambertian pattern [39]. Thus, the ceiling, walls and floor inside the data centre were modelled as Lambertian reflectors that have a reflection coefficient equal to 0.8 for the ceiling and walls and 0.3 for the floor [39]. Each surface consists of many small areas that act as a small second transmitters. They reflect the received signal in the form of a Lambertian pattern, where $n$ (emission order) is equal to 1. These small areas can have a significant impact on the results' time resolution. When these areas are very small, a higher resolution is obtained at the cost of longer simulation time. Therefore, each of these small areas was chosen to be 5 cm × 5 cm for the first-order reflection, and 20 cm × 20 cm for the second-order reflection to keep the simulation computation time within a reasonable limit [5], [27]. A communication floor (CF) was set at 0.25 m above the floor (see Figure 1a) which means that all communications are done above this CF. Below the CF rows of racks and racks can be connected through a cable.

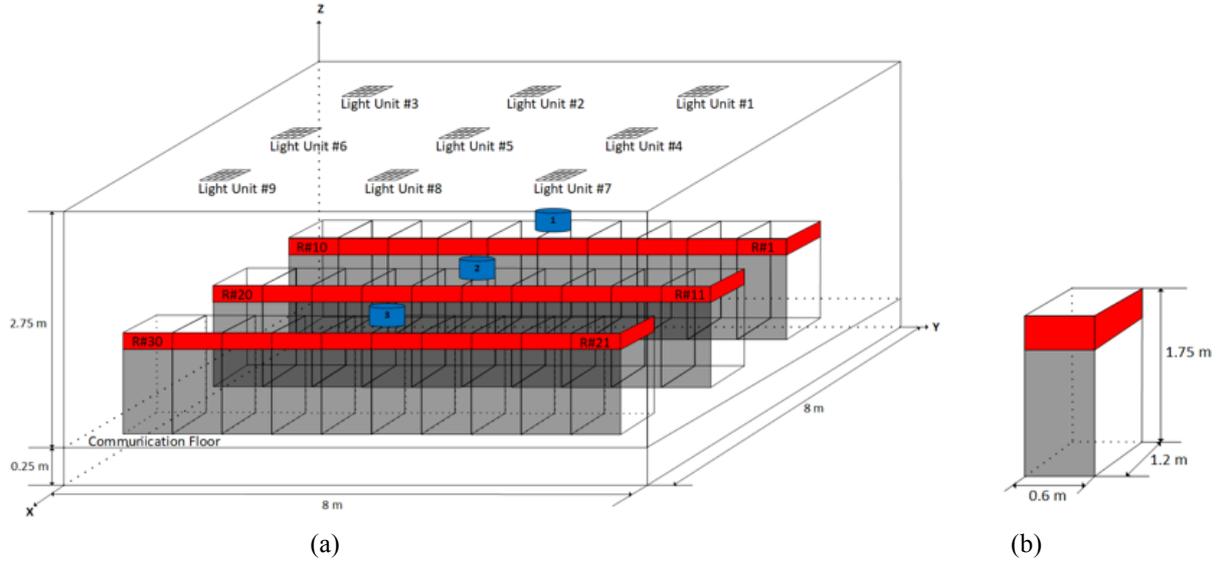

Figure 1: Optical wireless data centre system (a) Data Centre Configuration, (b) Rack diminutions

## 3. OPTICAL RECEIVER DESIGN

In this work, an optical receiver was placed on the top middle of each row of racks as shown in Figure 1a. Each receiver of each row of racks just covers the three transmitters that are placed 1 m directly above the row of racks. Two types of optical receiver were considered in this study. The first is an angle diversity receiver (ADR) that consists of three branches; branches 2 and 3 consist of three detectors while branch 1 contains 4 detectors as shown in Figure 2a. Branch 2 and 3 serve the six racks in the corner of each row, whereas branch 1 just serves the four racks in the middle of each row. The ADR was used to collect signals from a specific location while reducing interference. Each branch faces a different direction to cover different transmitters in the ceiling of the data centre by using the Azimuth ($Az$) and Elevation ($El$) angles. The $El$ angles of the three branches are set as follows: two branches have an angle equal to 25°, while the branch that faces upwards is set to 90°. The $Az$ angles of the branches are 0°, 90° and 270°. The FOV of each detector is set at 20°. In addition, each detector has an area equal to 20 $mm^2$. The second type of optical receiver is an imaging receiver (ImR) which contains 10 pixels for collecting signals and reducing interference. Four ImRs were located on the top middle of each row of racks. Each pixel has an area equal to 20 $mm^2$ with a narrow FOV. The ImR uses a lens that is placed above the pixels and the FOV of this lens equal to 65° to collect signals from the three transmitters above the row of racks.

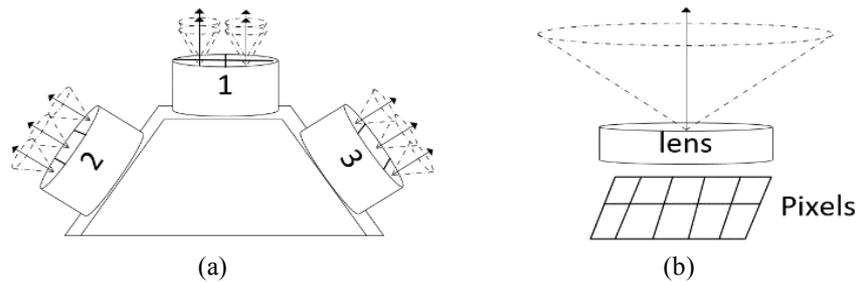

Figure 2: Optical Receiver Design: (a) ADR, (b) ImR.

## 4. OPTICAL TRANSMITTER DESIGN

In this study, RYGB LDs light units are used for illumination as well as a transmitter for communication. The data centre consists of nine RYGB Light units as shown in Figure 1a. Each one of these light units consists of 16 wide-semi angle RYGB LDs. The semi angle of each one of these RYGB LDs was set to 70⁰ to increase the illumination level inside the data centre. These light units were positioned in different fixed locations on the ceiling; (1.8 m, 2 m, 3 m), (1.8 m, 4 m, 3 m), (1.8 m, 6 m, 3 m), (4 m, 2 m, 3 m), (4 m, 4 m, 3 m), (4 m, 6 m, 3 m), (6.2 m, 2 m, 3 m), (6.2 m, 4 m, 3 m) and (6.2 m, 6 m, 3 m). The locations of these light units were chosen to offer good communication links for each row of racks. Thus, each row of racks collects signals from the three RYGB LDs light units that are placed 1 m above the row of racks.

## 5. SIMULATION SETUP AND RESULTS

In this work, each row of racks received the signal from the three RYGB LDs light units that are placed 1 m directly above the row of racks. In addition, each rack receives a wavelength different wavelength from its neighbour to provide a higher data rates. As mentioned above, a WDMA was used in this work. The assigned wavelength and light unit of each rack are shown in Table 1. The optical receiver that is placed in the top middle of each row of racks receives the desired wavelength from the selected light unit and transfers the received data to the router of the desired rack.

Table 1: The assigned wavelength and light unit of each rack

| Row 1 | | | Row 2 | | | Row 3 | | |
|---|---|---|---|---|---|---|---|---|
| Rack # | Wavelength | Light unit # | Rack # | Wavelength | Light unit # | Rack # | Wavelength | Light unit # |
| 1 | Red | 1 | 11 | Red | 4 | 21 | Red | 7 |
| 2 | Green | 1 | 12 | Green | 4 | 22 | Green | 7 |
| 3 | Yellow | 1 | 13 | Yellow | 4 | 23 | Yellow | 7 |
| 4 | Red | 2 | 14 | Red | 5 | 24 | Red | 8 |
| 5 | Green | 2 | 15 | Green | 5 | 25 | Green | 8 |
| 6 | Blue | 2 | 16 | Blue | 5 | 26 | Blue | 8 |
| 7 | Yellow | 2 | 17 | Yellow | 5 | 27 | Yellow | 8 |
| 8 | Yellow | 3 | 18 | Yellow | 6 | 28 | Yellow | 9 |
| 9 | Green | 3 | 19 | Green | 6 | 29 | Green | 9 |
| 10 | Red | 3 | 20 | Red | 6 | 30 | Red | 9 |

As outlined in the Section 3, two types of optical receivers are evaluated in this study; ADR and ImR. The SINR should be high to decrease the bit error rate (BER) which can be calculated when using on off keying (OOK) modulation as:

$$P_e = Q(\sqrt{SINR}) \quad (1)$$

where the function $Q(\cdot)$ is the Q-function. The SINR of each wavelength can be determined using:

$$SINR = \frac{R_w^2 \, (P_{s1w} - P_{s0w})^2}{\sigma_{tw}^2 + \sum_{l=1}^{L} R_w^2 (P_{i1w} - P_{i0w})^2} \quad (2)$$

where $R_w$ is the photodetector responsivity based on the used wavelength ($Red = 0.4 \frac{A}{W}$, $Yellow = 0.35 \frac{A}{W}$, $Green = 0.3 \frac{A}{W}$ and $Blue = 0.2 \frac{A}{W}$), $P_{s1w}$ is the optical power received from the desired wavelength that is associated with logic 1, $P_{s0w}$ is the received optical power from the desired wavelength that is associated with logic 0, $\sigma_{tw}$ is the total noise related to the received signal from the desired wavelength. $P_{i1w}$ is the received optical power from the other modulated light units using the same wavelength associated with logic 1 and $P_{i0w}$ is the optical power received from the other modulated light units using the same wavelength associated with logic 0.

The ImR consists of a lens as shown in Figure 2c with a transmission factor that depends on the incidence angle ($Y$) which is computed using [40]:

$$Tc(Y) = -0.1982Y^2 + 0.0425Y + 0.8778$$

Figure 3a compares the signal-to-interference-plus-noise ratio (SINR) (dB) of the two proposed optical receivers; the imaging and angle diversity receivers, using the selected combining method. The results show that both types of optical receivers provide high SINR, greater than 15.6 dB which is needed for $P_e = 10^{-9}$ [41]. The ImR offers higher SINR in many racks compared to the ADR. In terms of the supported data rate, the ImR can offer data rates up to 8.5 Gbps for each rack, while the supported data rate when using the ADR varied between 1.5Gbps and 7 Gbps, as shown in Figure 3b. These supported data rates are based on the SINR in Figure 3a. Thus, the ImR is better than ADR.

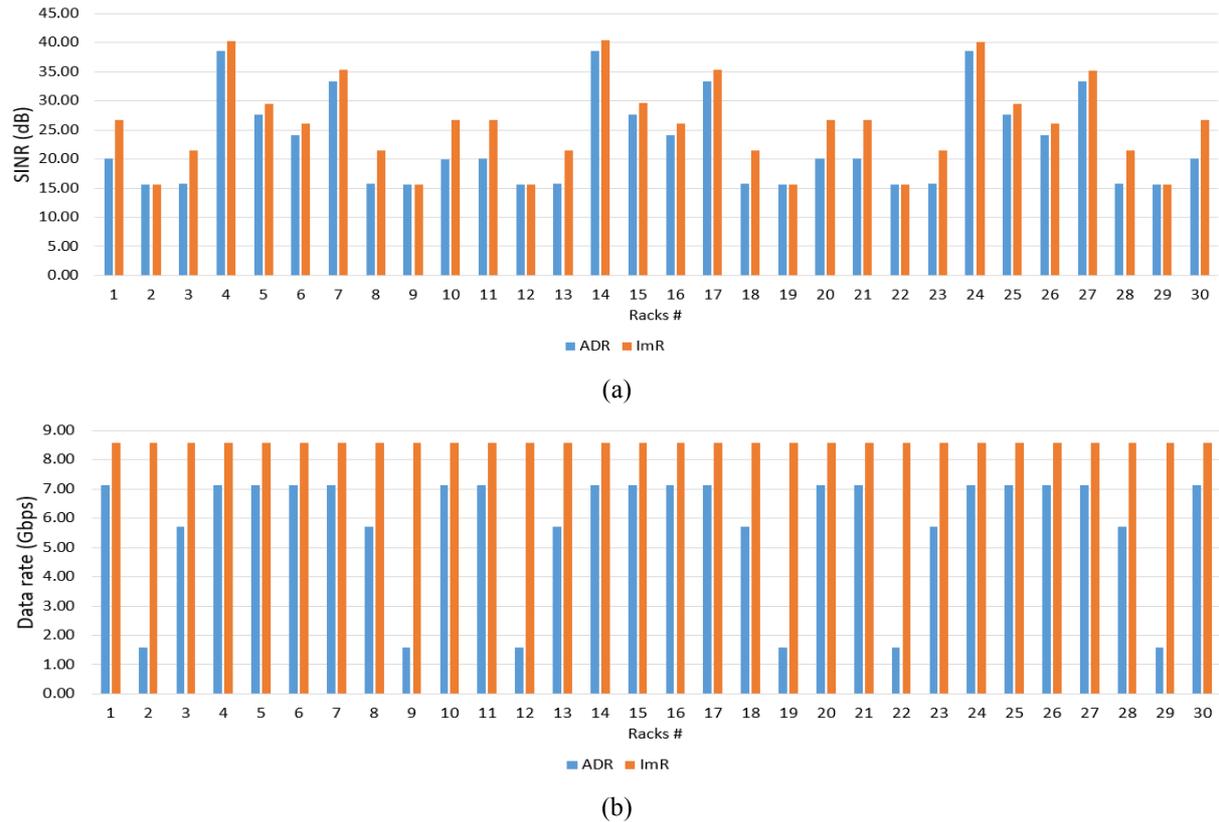

Figure 3: Optical wireless data centre system (a) Channel Bandwidth, (b) SINR and (c) Data Rate of ADR and ImR

## 6. CONCLUSIONS

In this paper, a data centre downlink was proposed using RYGB LDs as a transmitter. The proposed data centre contains 30 racks that are divided into three rows. Each row consists of 10 racks and each rack has a top of rack switch that is placed on the top of the rack. Two types of optical receivers were examined in the proposed data centre; an angle diversity receiver (ADR) and an imaging receiver (ImR), using a simple modulation technique, OOK. A data rate of 8.5 Gbps can be achieved for downlink communication for each row at the same time by using the ImR, while the ADR provides varied data rates between 1.5Gbps and 7 Gbps. The ImR provides better results compared to the ADR in terms of the supported data rate.

## ACKNOWLEDGEMENTS

The authors would like to acknowledge funding from the Engineering and Physical Sciences Research Council (EPSRC) INTERNET (EP/H040536/1), STAR (EP/K016873/1) and TOWS (EP/S016570/1) projects. The authors extend their appreciation to the deanship of Scientific Research under the International Scientific Partnership Program ISPP at King Saud University, Kingdom of Saudi Arabia for funding this research work through ISPP#0093. OZA would like to thank Umm Al-Qura University in the Kingdom of Saudi Arabia for funding his PhD scholarship. SOMS would like to thank the University of Leeds and the Higher Education Ministry in Sudan for funding her PhD scholarship. SHM would like to thank EPSRC for providing her Doctoral Training Award scholarship. All data are provided in full in the results section of this paper.

# REFERENCES

1. Z. Ghassemlooy, W. Popoola, and S. Rajbhandari, *Optical wireless communications: system and channel modelling with Matlab®*. 2012.
2. F. E. Alsaadi, M. A. Alhartomi, and J. M. H. Elmirghani, "Fast and efficient adaptation algorithms for multi-gigabit wireless infrared systems," *J. Light. Technol.*, vol. 31, no. 23, pp. 3735–3751, 2013.
3. A. T. Hussein and J. M. H. Elmirghani, "10 Gbps Mobile Visible Light Communication System Employing Angle Diversity, Imaging Receivers, and Relay Nodes," *J. Opt. Commun. Netw.*, vol. 7, no. 8, pp. 718–735, 2015.
4. S. H. Younus and J. M. H. Elmirghani, "WDM for high-speed indoor visible light communication system," in *International Conference on Transparent Optical Networks*, 2017, pp. 1–6.
5. A. T. Hussein, M. T. Alresheedi, and J. M. H. Elmirghani, "20 Gb/s Mobile Indoor Visible Light Communication System Employing Beam Steering and Computer Generated Holograms," *J. Light. Technol.*, vol. 33, no. 24, pp. 5242–5260, 2015.
6. A. T. Hussein, M. T. Alresheedi, and J. M. H. Elmirghani, "25 Gbps mobile visible light communication system employing fast adaptation techniques," in *2016 18th International Conference on Transparent Optical Networks (ICTON)*, 2016.
7. O. Z. Alsulami, M. T. Alresheedi, and J. M. H. Elmirghani, "Transmitter diversity with beam steering," in *2019 21st International Conference on Transparent Optical Networks (ICTON)*, 2019, pp. 1–5.
8. O. Z. Alsulami, M. O. I. Musa, M. T. Alresheedi, and J. M. H. Elmirghani, "Visible light optical data centre links," in *2019 21st International Conference on Transparent Optical Networks (ICTON)*, 2019, pp. 1–5.
9. Cisco Mobile, *Cisco Visual Networking Index: Global Mobile Data Traffic Forecast Update, 2016-2021 White Paper*. 2017.
10. K. L. Sterckx, J. M. H. Elmirghani, and R. A. Cryan, "Sensitivity assessment of a three-segment pyrimadal fly-eye detector in a semi-disperse optical wireless communication link," *IEE Proc. Optoelectron.*, vol. 147, no. 4, pp. 286–294, 2000.
11. A. G. Al-Ghamdi and J. M. H. Elmirghani, "Performance evaluation of a triangular pyramidal fly-eye diversity detector for optical wireless communications," *IEEE Commun. Mag.*, vol. 41, no. 3, pp. 80–86, 2003.
12. M. T. Alresheedi and J. M. H. Elmirghani, "Hologram Selection in Realistic Indoor Optical Wireless Systems With Angle Diversity Receivers," *IEEE/OSA J. Opt. Commun. Netw.*, vol. 7, no. 8, pp. 797–813, 2015.
13. A. T. Hussein, M. T. Alresheedi, and J. M. H. Elmirghani, "Fast and Efficient Adaptation Techniques for Visible Light Communication Systems," *J. Opt. Commun. Netw.*, vol. 8, no. 6, pp. 382–397, 2016.
14. S. H. Younus, A. A. Al-Hameed, A. T. Hussein, M. T. Alresheedi, and J. M. H. Elmirghani, "Parallel Data Transmission in Indoor Visible Light Communication Systems," *IEEE Access*, vol. 7, pp. 1126–1138, 2019.
15. A. G. Al-Ghamdi and M. H. Elmirghani, "Optimization of a triangular PFDR antenna in a fully diffuse OW system influenced by background noise and multipath propagation," *IEEE Trans. Commun.*, vol. 51, no. 12, pp. 2103–2114, 2003.
16. A. G. Al-Ghamdi and J. M. H. Elmirghani, "Characterization of mobile spot diffusing optical wireless systems with diversity receiver," in *ICC'04 IEEE International Conference on Communications*, 2004.
17. F. E. Alsaadi and J. M. H. Elmirghani, "Performance evaluation of 2.5 Gbit/s and 5 Gbit/s optical wireless systems employing a two dimensional adaptive beam clustering method and imaging diversity detection," *IEEE J. Sel. Areas Commun.*, vol. 27, no. 8, pp. 1507–1519, 2009.
18. O. Z. Alsulami, M. T. Alresheedi, and J. M. H. Elmirghani, "Optical Wireless Cabin Communication System," in *2019 IEEE Conference on Standards for Communications and Networking (CSCN)*, 2019, pp. 1–4.
19. O. Z. Alsulami, M. O. I. Musa, M. T. Alresheedi, and J. M. H. Elmirghani, "Co-existence of Micro, Pico and Atto Cells in Optical Wireless Communication," in *2019 IEEE Conference on Standards for Communications and Networking (CSCN)*, 2019, pp. 1–5.
20. S. O. M. Saeed, S. Hamid Mohamed, O. Z. Alsulami, M. T. Alresheedi, and J. M. H. Elmirghani, "Optimized resource allocation in multi-user WDM VLC systems," in *2019 21st International Conference on Transparent Optical Networks (ICTON)*, 2019, pp. 1–5.
21. M. T. Alresheedi and J. M. H. Elmirghani, "10 Gb/s indoor optical wireless systems employing beam delay, power, and angle adaptation methods with imaging detection," *IEEE/OSA J. Light. Technol.*, vol. 30, no. 12, pp. 1843–1856, 2012.
22. K. L. Sterckx, J. M. H. Elmirghani, and R. A. Cryan, "Pyramidal fly-eye detection antenna for optical wireless systems," *Opt. Wirel. Commun. (Ref. No. 1999/128), IEE Colloq.*, pp. 5/1-5/6, 1999.
23. F. E. Alsaadi, M. Nikkar, and J. M. H. Elmirghani, "Adaptive mobile optical wireless systems employing a beam clustering method, diversity detection, and relay nodes," *IEEE Trans. Commun.*, vol. 58, no. 3, pp. 869–879, 2010.
24. F. E. Alsaadi and J. M. H. Elmirghani, "Adaptive mobile line strip multibeam MC-CDMA optical wireless system employing imaging detection in a real indoor environment," *IEEE J. Sel. Areas Commun.*, vol. 27, no. 9, pp. 1663–1675, 2009.
25. M. T. Alresheedi and J. M. H. Elmirghani, "Performance evaluation of 5 Gbit/s and 10 Gbit/s mobile optical wireless systems employing beam angle and power adaptation with diversity receivers," *IEEE J. Sel. Areas Commun.*, vol. 29, no. 6, pp. 1328–1340, 2011.
26. F. E. Alsaadi and J. M. H. Elmirghani, "Mobile Multi-gigabit Indoor Optical Wireless Systems Employing Multibeam Power Adaptation and Imaging Diversity Receivers," *IEEE/OSA J. Opt. Commun. Netw.*, vol. 3, no. 1, pp. 27–39, 2011.
27. A. G. Al-Ghamdi and J. M. H. Elmirghani, "Line Strip Spot-Diffusing Transmitter Configuration for Optical Wireless Systems Influenced by Background Noise and Multipath Dispersion," *IEEE Trans. Commun.*, vol. 52, no. 1, pp. 37–45, 2004.
28. F. E. Alsaadi and J. M. H. Elmirghani, "High-speed spot diffusing mobile optical wireless system employing beam angle and power adaptation and imaging receivers," *J. Light. Technol.*, vol. 28, no. 16, pp. 2191–2206, 2010.
29. G. Cossu, a M. Khalid, P. Choudhury, R. Corsini, and E. Ciaramella, "3.4 Gbit/s visible optical wireless transmission based on RGB LED.," *Opt. Express*, vol. 20, no. 26, pp. B501-6, 2012.
30. Y. Wang, Y. Wang, N. Chi, J. Yu, and H. Shang, "Demonstration of 575-Mb/s downlink and 225-Mb/s uplink bi-directional SCM-WDM visible light communication using RGB LED and phosphor-based LED," *Opt. Express*, vol. 21, no. 1, p. 1203, 2013.
31. A. Neumann, J. J. Wierer, W. Davis, Y. Ohno, S. R. J. Brueck, and J. Y. Tsao, "Four-color laser white illuminant demonstrating high color-rendering quality," *Opt. Express*, vol. 19, no. S4, p. A982, 2011.
32. F.-M. Wu, C.-T. Lin, C.-C. Wei, C.-W. Chen, Z.-Y. Chen, and K. Huang, "3.22-Gb/s WDM Visible Light Communication of a Single RGB LED Employing Carrier-Less Amplitude and Phase Modulation," in *Optical Fiber Communication Conference/National Fiber Optic Engineers Conference 2013*, 2013, p. OTh1G.4.
33. T. A. Khan, M. Tahir, and A. Usman, "Visible light communication using wavelength division multiplexing for smart spaces," *Consumer Communications and Networking Conference (CCNC), 2012 IEEE*. pp. 230–234, 2012.
34. J.M.H. Elmirghani, T. Klein, K. Hinton, L. Nonde, A.Q. Lawey, T.E.H. El-Gorashi, M.O.I. Musa, and X. Dong, "GreenTouch GreenMeter Core Network Energy Efficiency Improvement Measures and Optimization [Invited]," IEEE/OSA Journal of Optical Communications and Networking, vol. 10, No. 2, pp. 250-269, 2018.
35. L. Nonde, T. E. H. El-Gorashi, and J. M. H. Elmirghani, "Energy Efficient Virtual Network Embedding for Cloud Networks," *J. Light. Technol.*, vol. 33, no. 9, pp. 1828–1849, 2015.
36. H. M. M. Ali, T. E. H. El-Gorashi, A. Q. Lawey, and J. M. H. Elmirghani, "Future Energy Efficient Data Centers with Disaggregated Servers," *J. Light. Technol.*, vol. 35, no. 24, pp. 5361–5380, 2017.
37. A. M. Al-Salim, A. Q. Lawey, T. E. H. El-Gorashi, and J. M. H. Elmirghani, "Energy Efficient Big Data Networks: Impact of Volume and Variety," *IEEE Trans. Netw. Serv. Manag.*, vol. 15, no. 1, pp. 458–474, 2018.
38. J. R. Barry, J. M. Kahn, W. J. Krause, E. A. Lee, and D. G. Messerschmitt, "Simulation of Multipath Impulse Response for Indoor Wireless Optical Channels," *IEEE J. Sel. Areas Commun.*, vol. 11, no. 3, pp. 367–379, 1993.
39. F. R. Gfeller and U. Bapst, "Wireless In-House Data Communication via Diffuse Infrared Radiation," *Proc. IEEE*, vol. 67, no. 11, pp. 1474–1486, 1979.

40. P. Djahani and J. M. Kahn, "Analysis of infrared wireless links employing multibeam transmitters and imaging diversity receivers," *IEEE Trans. Commun.*, vol. 48, no. 12, pp. 2077–2088, 2000.
41. M.T. Alresheedi, A. T. Hussein, and J.M.H. Elmirghani, "Uplink Design in VLC Systems with IR Sources and Beam Steering," IET Communications, vol. 11, No. 3, pp. 311-317, 2017.